\documentclass[]{raa}            

\usepackage{graphicx,times}
\usepackage{natbib}
\usepackage{amssymb,amsmath}
\bibpunct{(}{)}{;}{a}{}{,}           
\usepackage{longtable}
\usepackage[pagebackref=true]{hyperref}

\def\cm2{cm$^{-2}$}

\def\nh3{NH$_3$}
\def\n2h{N$_2$H$^+$}

\def\13co{$^{13}$CO}
\def\c18o{C$^{18}$O}
\def\hc3n{HC$_3$N}
\def\h2{H$_2$}
\def\nh{n(H$_2$)}

\begin{document}
  \title{The FAST Discovery of a Millisecond Pulsar M15O (PSR J2129+1210O) Hidden in the Harmonics of M15A (PSR J2129+1210A)}
   \volnopage{Vol.0 (20xx) No.0, 000--000}     
   \setcounter{page}{1}          
   \author{Yinfeng Dai 
      \inst{1}
   \and Zhichen Pan
      \inst{2,3,4,5}
   \and Lei Qian
      \inst{2,3,4,5}
   \and Liyun Zhang
      \inst{1,6}
   \and Dejiang Yin
      \inst{1}
   \and Baoda Li
      \inst{1}
   \and Yaowei Li 
      \inst{1} 
   \and Yuxiao Wu 
      \inst{7}   
   \and Yujie Lian 
      \inst{8,9} 
   }

   \institute{College of Physics, Guizhou University, Guiyang 550025, China;{\it liy\_zhang@hotmail.com}\\
        \and
             National Astronomical Observatories, Chinese Academy of Sciences,
             Beijing 100010, China; {\it panzc@nao.cas.cn;lqian@nao.cas.cn}\\
        \and
             Guizhou Radio Astronomical Observatory, Guizhou University,
             Guiyang 550025, China\\
        \and
             College of Astronomy and Space Sciences, University of Chinese Academy of Sciences,
             Beijing 100049, China\\
        \and
             Key Laboratory of Radio Astronomy, Chinese Academy of Sciences,
             Beijing 100101, China\\
        \and     
             International Centre of Supernovae, Yunnan Key Laboratory, Kunming 650216, China\\
        \and
             School of Science, Chongqing University of Posts and Telecommunications, Chongqing 40000, China\\
       \and     
             Institute for Frontiers in Astronomy and Astrophysics, Beijing Normal University, Beijing 102206, China\\
        \and     
             School of Physics and Astronomy, Beijing Normal University, Beijing 100875, China\\
   }

   \date{Received~~2025 March 24; accepted~~2025 April 20}

\abstract{We report the discovery of an isolated millisecond pulsar M15O (J2129+1210O) from the globular cluster M15 (NGC 7078) with a period of $\sim$11.06686 ms and a dispersion measure of $\sim$67.44 cm$^{-3}$ pc. Its spin period is so close to the 10th harmonic of the bright pulsar M15A ($\sim$11.06647 ms) that it was missed in previous pulsar search. We suggest adding the spectrum in the pulsar candidate diagnostic plot to identify new signals near the harmonics. M15O has the first spin frequency derivative and the second spin frequency derivative,being 1.79191(5) $\times$ $10^{-14}$ Hz $s^{-1}$ and 3.3133(6)$\times$ $10^{-23}$ Hz $s^{-2}$, respectively. Its projected distance from the optical center of M15 is the closest among all the pulsars in M15. The origin can be something from the center of the massive and core-collapsed globular cluster M15.
\keywords{methods: data analysis - globular clusters: individual (M15) - pulsars: individual (M15O or J2129-1210O)} 
}
   \authorrunning{Dai et al.}            
   \titlerunning{The FAST Discovery of a Millisecond Pulsar in M15 }  
   \maketitle

%

\section{Introduction} 
\label{sect:intro}
Pulsars are rotating neutron stars with strong dipole magnetic fields, 
firstly discovered in 1967 through radio observations \citep{1968Natur.217..709H}. 
The discoveries of pulsars provide a natural laboratory for extreme physical environments,
such as binary pulsars (e.g., PSR B1913+16;\citealt{1975ApJ...195L..51H}), 
massive neutron stars (e.g., PSR J1614$-$2230;\citealt{2010Natur.467.1081D}), and millisecond pulsars (MSPs; e.g., PSR B1937+21 ;\citealt{1982Natur.300..615B}).
Pulsar searches face challenges from increasing radio frequency interference (RFI),
dispersion smearing due to free electrons in the interstellar medium, 
and signal scattering caused by irregularities in interstellar plasma \citep{2006MsT..........5B}.
If the pulsars are in binary systems, 
the orbital motion and rotational drift of pulsars will affect the times of arrival(ToAs) of the pulses to reach the Earth, 
significantly increasing the detection complexity of pulsar signals (e.g., \citealt{1990Natur.346...42A}).

The prevalent pulsar search methods mainly utilize the Fourier domain acceleration search 
(FFT; e.g., \textsc{PRESTO}\footnote{\url{https://github.com/scottransom/presto}} software suite; \citealt{2002AJ....124.1788R})
and the Fast Folding Algorithm search (e.g., RIPTIDE-FFA\footnote{\url{https://riptide-ffa.readthedocs.io/en/latest/index.html}};\citealt{2020MNRAS.497.4654M}. 
Both of these methods necessitate period folding of time domain files and subsequent signal identification through folded profiles \citep{2002AJ....124.1788R}.  
Cross-checking with multiple observations helps to verify candidates, to either confirm or deny a candidate.
In addition, the stack search, which involves incoherently stacking power spectra across multiple observations, 
has been proven effective in enhancing the detection sensitivity of faint pulsars using archival data 
(e.g., \citealt{1993PhDT.........2A}; \citealt{2016MNRAS.459L..26P};  \citealt{2018ApJ...855..125C}).

Globular clusters (GCs) are characterized by the high stellar densities in their cores and frequent dynamical interactions, 
such as exchange interactions and tidal trapping (\citealt{1996AJ....112.1487H}, 2010 edition).
GCs have become prolific sites for the discovery of MSPs (e.g., \citealt{2014ApJ...795...29P}).
Pulsars in GCs are most prominently concentrated in compact spatial distributions, 
such as those in Terzan 5 (e.g., \citealp{2005Sci...307..892R};\citealp{2024A&A...686A.166P}) and 47 Tucanae (NGC~104; e.g., \citealp{2017MNRAS.471..857F}),
well within the half-power beamwidth of one beam for single-dish telescopes at L-band (e.g., $\sim 3'$).
In this environment, detection and differentiation of pulsar signals become somehow difficult with multiple pulsars especially with very bright pulsars as their signals may interfere with each other.
For GCs hosting many known pulsars with similar dispersion measure(DM) values, this increases the difficulty of distinguishing a potential pulsar signal.
Sometimes, the timing analysis has to be used to confirm a new pulsar as their spin periods can be very close (e.g., M3D and M3E; \citealp{2024ApJ...972...43L}).

To determine whether a detected signal in the same direction corresponds to a known pulsar, 
we can compare the pulsar's spin period and DM values. 
If there are pulsars with strong fluxes, the candidate list may contain numerous instances of their harmonics, especially for pulsars with long spin periods. 
If the period of a new pulsar is similar to or overlaps with that of a known pulsar or its harmonics, 
and their DM values are also close (owing to the concentrated DM distribution within the cluster),
conventional folding algorithms struggle to separate the signals due to phase - alignment ambiguities. 
The standard screening process may fail to effectively distinguish between them, 
potentially leading to missing the new pulsar.

The Five-hundred-meter Aperture Spherical Radio Telescope (FAST, \citealt{2011IJMPD..20..989N}) operates with the most sensitivity among all the single-dish radio telescopes due to its illuminated aperture of 300 m.
There are 45 GCs in the FAST sky, and pulsars have been discovered in 17 of them. 
Among them, M15 (NGC~7078) has 14 pulsars ranked second after NGC~6517.
It is one of the oldest core-collapsed clusters with a core density of $\rho_0 = 10^{5.05} \, \rm L_{\odot} \, {\rm pc}^{-3}$ and with a distance from the Sun of 10.4 kpc (\citealt{1996AJ....112.1487H}, 2010 edition).
A simulation predicts that it has the highest number of predicted pulsars among all the GCs in FAST sky \citep{2024ApJ...969L...7Y}.
Therefore, M15 has become a high-priority target for searching pulsars.

In this study, we report the discovery of a new pulsar (M15O, J2129-1210O) by employing FFT and the stack search method.
It hides near the 10th harmonic of PSR J2129+1210A (M15A), identified by visual inspection at the spectra directly, 
and was finally verified by timing analysis using 
\textsc{TEMPO}\footnote{\url{http://nanograv.github.io/tempo/}} \citep{2015ascl.soft09002N}.
The subsequent Section \ref{data} is for the details of the data and data reduction procedures.
Section \ref{results} describes the discovery and timing results of M15O.
Section \ref{discussion} presents the discussion regarding this new pulsar, and the conclusions provided in Section \ref{conclusion}.


\section{Data and Data Reduction}
\label{data}

This study is based on the 19 observations of M15 conducted by FAST between November 2019 and February 2024. 
The longest single observation lasted up to 4.5 hr.
The total integration time was $\sim$44 hr.
A detailed summary of these archival data is provided in Table~\ref{Tab1}. 
These data were recorded by the central beam of the FAST 19-beam receiver in the L-band with 4096 frequency channels (each with a bandwidth of 0.122 MHz) and 
it covers the frequency range from 1.05 to 1.45 GHz \citep{2019SCPMA..6259502J}.

With 3$'$ beam size, all known pulsars in M15 can be observed in at least some of these data.
We reprocessed these data with FFT and the stack search methods, aiming at searching for new pulsars in M15.
In the latter method, the stacked power spectra at different DMs are obtained by incoherently summing the power spectra of all observations.

\begin{table}
\centering
\setlength{\tabcolsep}{4pt} 
\caption{FAST observation data for M15 used in this work and the detected results of M15O.
Most observations were made in tracking mode, 
except for the observation on 2022 January 2, 
which was conducted in snapshot mode using only the first quarter of the time. 
The symbol ``$\circ$'' indicates that the signal of M15O was detected in the \textsc{PRESTO} search results, the symbol ``$\surd$'' means the data were used for timing analysis, and the ``$\star$'' symbol signifies that the observation is used in the stack search method.
}
\begin{tabular}{cccccccc}
  \hline\hline
  \multicolumn{1}{c}{Observation Date} & \multicolumn{1}{c}{MJD} & \multicolumn{1}{c}{$T_{\rm obs}$\,(s)} & \multicolumn{1}{c}{Stack} & \multicolumn{1}{c}{FFT} & \multicolumn{1}{c}{Timing} & \multicolumn{1}{c}{SNR} & \multicolumn{1}{c}{Flux Density ($\mu\mathrm{Jy}$)} \\
  \hline
    2019.11.09 &  58796  & 4800  & $\star$ &  & $\surd$ & 8.17 & 2.19 \\
    2020.08.30 & 59091  & 3300  & $\star$ &  &  &  & \\
    2020.09.22 & 59114 &  9000   & $\star$ & & $\surd$ & 6.60 & 1.29\\
    2020.09.25 & 59117&  14400  & $\star$ & $\circ$ & $\surd$ & 12.18 & 1.88 \\
    2020.12.21 &  59204  & 16200  & $\star$ & & $\surd$ & 9.18  & 1.34\\
    2021.03.09 &  59282 & 8700   & $\star$ & $\circ$ & $\surd$ & 8.56 & 1.70\\
    2022.01.02 &  59581   & 1800  & $\star$ & & $\surd$ & 5.21 & 2.28\\
    2022.09.04 &  59826 & 1800  & $\star$ &  &  &  &  \\
    2022.10.15 & 59867 &  3000  & $\star$ &  & &   &   \\
    2022.11.19 &  59902   &6600   & $\star$ & $\circ$ & $\surd$ & 11.89 & 2.72 \\
    2022.12.20 &  59933  & 10200   & $\star$ & $\circ$ & $\surd$ & 9.35 & 1.72 \\
    2023.01.20 &  59964  & 10200 & $\star$ &  & $\surd$ & 7.83 & 1.44\\
    2023.02.20 & 59995  &  6600   & $\star$ &  & $\surd$ & 4.65  & 1.06 \\
    2023.12.06 &  60284  & 13560   & $\star$ &  & $\surd$ & 6.44   & 1.03\\
    2023.12.19 &  60297  & 15840   & $\star$ & $\circ$ & $\surd$ & 9.47  & 1.40\\
    2024.01.22 &  60331  & 9060  & $\star$ &  & $\surd$  & 5.25  & 1.02 \\
    2024.01.23 &  60332 & 10800   & $\star$ &  & $\surd$ & 6.44 & 1.15 \\
    2024.01.30 &  60339  & 6320  & $\star$ &  & $\surd$ & 6.91  & 1.61 \\
    2024.02.17 & 60357  &  6375   & $\star$ &  & $\surd$  & 7.65 & 1.78 \\
  \hline
  Total &  &  & 19 & 5 & 16 \\ 
  \hline
\end{tabular}
\label{Tab1}
\end{table}

\subsection{Fourier Domain Acceleration Search}

The RFI in the data was labeled by applying the \texttt{rfifind} routine of \textsc{PRESTO}.  
The DM values of the known pulsars in M15 are distributed in the range 65.5 to 67.8 \,cm$^{-3}$\,pc, 
with the mean value approximated as 67.0 \,cm$^{-3}$\,pc.
The \texttt{prepsubband} routine was used to dedisperse the PSRFITS data over a larger DM trial range from 63 to 70 \,cm$^{-3}$\,pc, with step size of 0.05 \,cm$^{-3}$\,pc.
In order to reduce the interference caused by low-frequency noise, the \texttt{rednoise} routine was applied.
The \texttt{realfft} routine was utilized to transform the dedispersed time series to the frequency domain.
The \texttt{accelsearch} routine was used to conduct acceleration searches in the spectra. 
The isolated pulsars dominate the pulsar population in M15 (13 out of 14). 
Thus, a low $z_{\rm max}$ value of 20 was selected ($z_{\rm max}$ means the largest Doppler drift of the pulse frequency; \citealt{2002AJ....124.1788R}). 
Finally, an updated version of \texttt{JinglePulsar\footnote{\url{https://github.com/jinglepulsar/jinglesifting}}} \citep{2021RAA....21..143P} was applied to sift and fold promising candidates for visual inspection.

\begin{figure}
\centering
\includegraphics[width=\textwidth, angle=0]{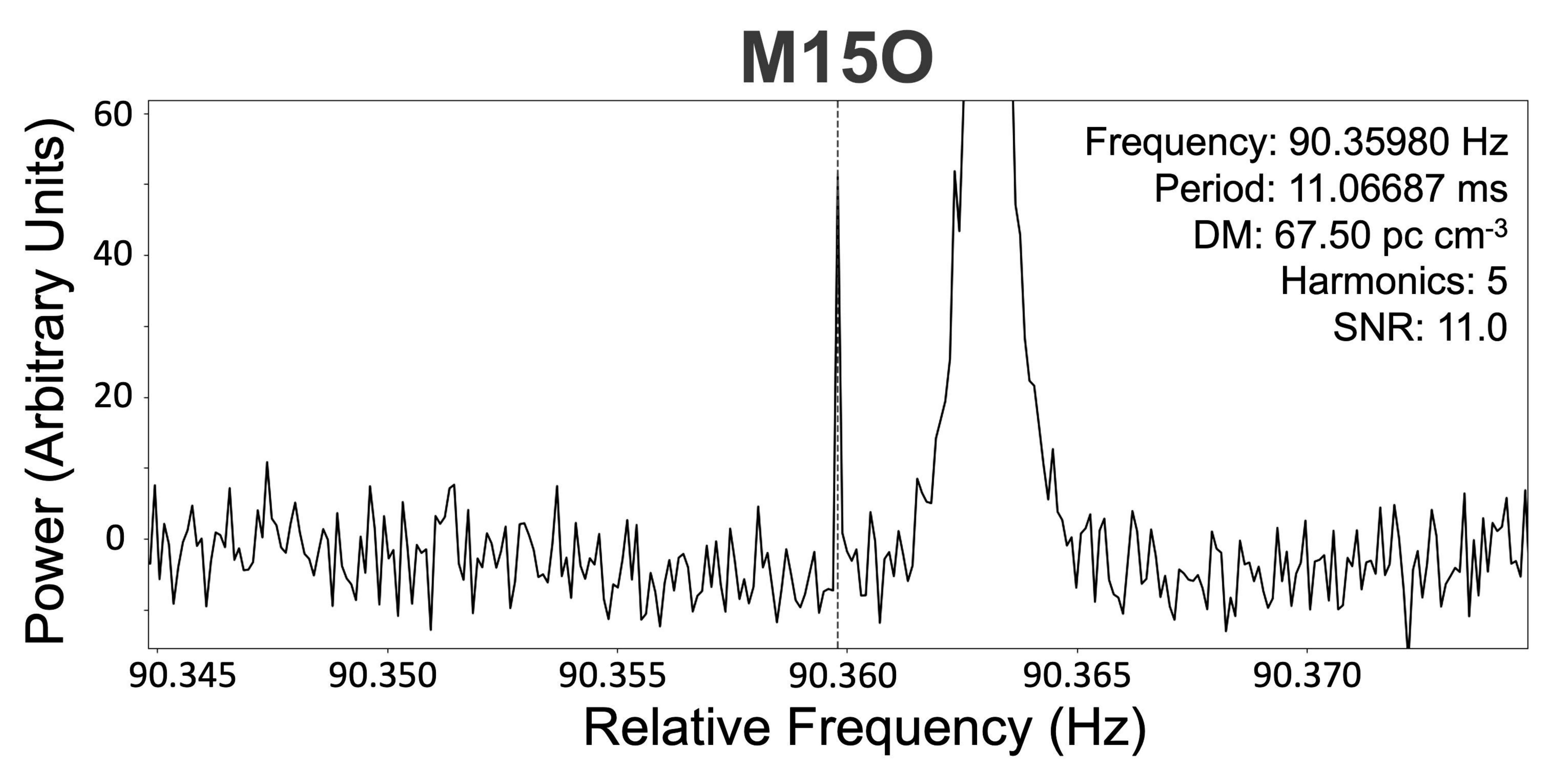}
\caption{The power spectrum around the frequency of M15O from DM value of 67.50 \,cm$^{-3}$\,pc.
There is a strong signal next to the frequency of M15O ($\sim$90.35980 Hz). 
This strong signal's frequency is $\sim$90.36306 Hz, 
which is the 10th harmonic of the signal of M15A \citep{2024ApJ...974L..23W}.
To avoid effects from the M15A harmonics and to get an accurate SNR of this signal, 
only the part on the left side of the signal was taken as noise in the power spectrum. 
When the number of harmonic stacking is 5, SNR has a maximum value of $\sim$11.0.}
\label{Fig1}
\end{figure}
\begin{figure}
\centering
\includegraphics[width=\textwidth, angle=0]{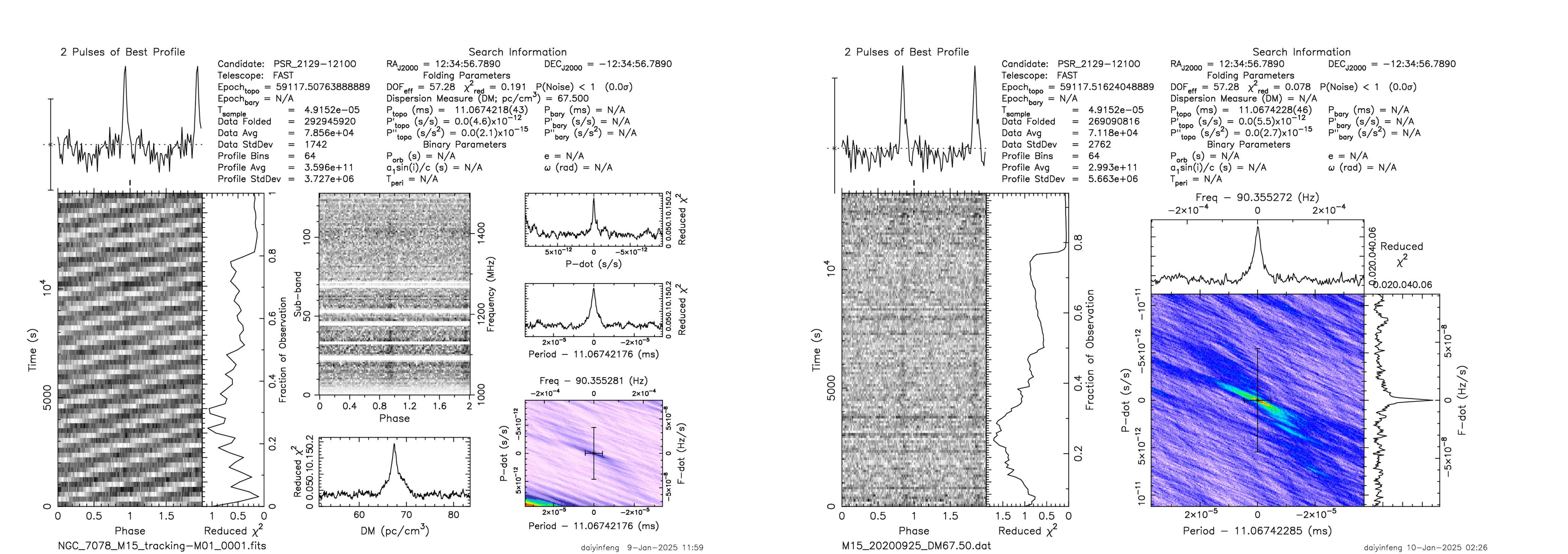}
\caption{The optimal detection plots of M15O (on MJD 59117). 
They were generated using the \texttt{prepfold} routine from \textsc{PRESTO}.
The left plot was generated using the PSRFITS data, and the right plot was generated using the time series obtained by chopping the  M15A signal with the \texttt{simple\_zapbirds.py} routine of \textsc{PRESTO}. 
}
\label{Fig2}
\end{figure}

\subsection{The Stack Search}

The stack search also firstly was performed with the routines \texttt{rfifind}, \texttt{prepsubband}, and \texttt{realfft} in \textsc{PRESTO} in the same steps as the above FFT.
With these steps, a total of 141 frequency domain series files can be obtained from different DM values.
From each file, the real and imaginary parts are added in an orthogonal manner to form the power values.
Power spectra from multiple datasets with the same DM values are added to improve the signal-to-noise ratio (SNR) of possible periodic candidates.
We added the power of the harmonics (up to 16)
sequentially to the fundamental frequency in order to mitigate the energy dispersion in the frequency domain.
After each harmonic addition, the power spectrum was retained for subsequent analysis.
Finally, signals with power values exceeding $5 {\sigma}$ ($\sigma$ is the standard deviation of the power value) in each stacked power spectrum are extracted as candidates.
These candidates were all drawn on the power spectrum, and then screened one by one by visual inspection.
Moreover the optimal DM trial and spin frequency of each 
candidate were further used to fold.
More details about the stack search method will be described in Dai et al. (2025 in preparation).

\section{Results}
\label{results}

We discovered three pulsars in total, namely PSR J2129+1210M, N and O (M15M, N, and O) with the FFT and stack search schemes.
As the M15M and N were firstly discovered by the stacking method, we will mention them in the next paper (Dai et al. 2025, in preparation).
In this paper, we focus on the discovery of M15O and the reason for missing it in previous searches.

With the FFT method, we detected all know pulsars in M15.
In addition, a large number of high-frequency harmonics appear in the candidates, being harmonics mainly from M15A and M15B.
By comparing their period and DM, it is difficult to distinguish whether two signals with similar periods are harmonics from known pulsars or new pulsars.
Multiple pulsar signals can only be distinguished by precise timing measurements, especially in dense clusters (e.g., Terzan 5; \citealp{2005Sci...307..892R}).
Eventually, a pulsar named M15O (J2129-1210O) with a period of $\sim$11.06687 ms was detected.
M15O stands out with a relatively high SNR of $\sim$11.0.
Its spin period is very close to the 10th harmonic of M15A ($\sim$11.06647ms),  
with a difference of only $\sim$0.004 ms.
With the 10th harmonic of M15A, the power spectrum of this pulsar is displayed in Figure ~\ref{Fig1}.

The folded plot of M15O using the \textsc{PRESTO} \texttt{prepfold} routine is shown in Figure ~\ref{Fig2}.
The anomalous morphology in the time domain depicted in Figure~\ref{Fig2} (left panel) arises from phase interference between M15O and harmonics of M15A.
The diagonal striations in the time series are from the M15A's harmonics.
To further validate our findings, we chopped the M15A's signal in the frequency domain using \textsc{PRESTO}'s \texttt{simple\_zapbirds.py}. 
Then, a time domain series was generated by the inverse Fourier transformation using \texttt{realfft}.
After folding that time series, the M15O signal remains, yet without the harmonic of M15A.

\begin{figure}
\centering
\includegraphics[width=15cm]{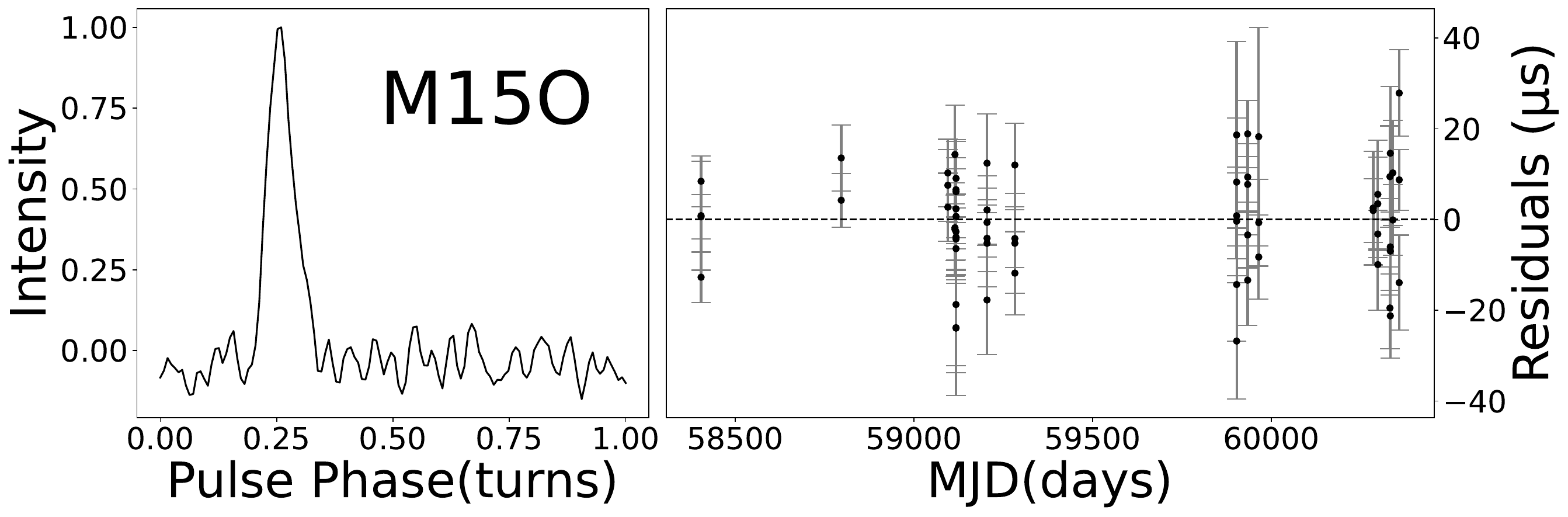}
\caption{The average pulse profile (left) and the timing residual (right) of M15O. 
The integrated pulse profile was obtained by summing all the detections over 64 pulse phase bins.
}
\label{Fig3}
\end{figure}

\begin{figure}
\centering
\includegraphics[width=15cm]{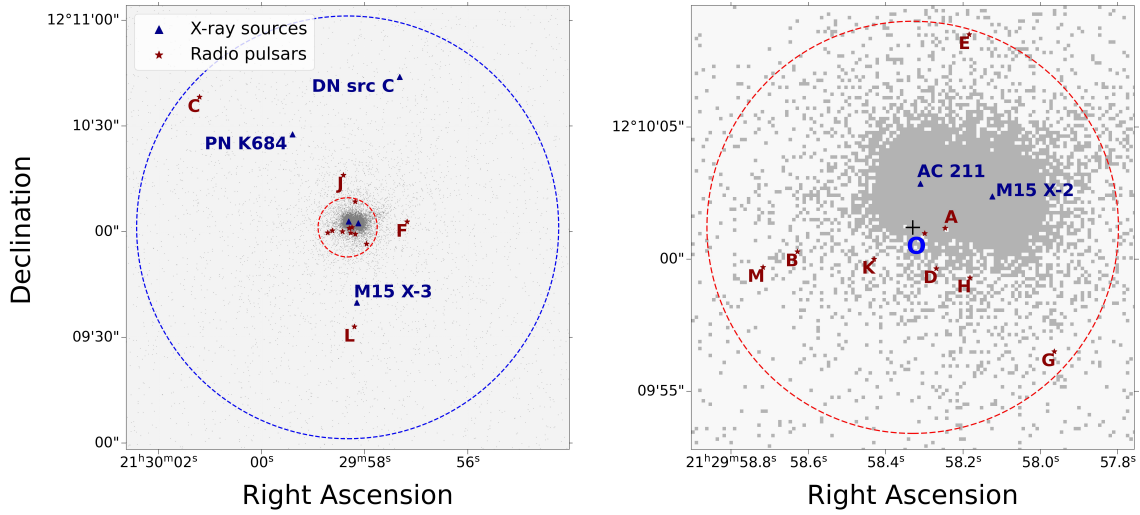}
\caption{
The background is the image of M15 {\it Chandra} X-ray archival data (OBsID 2413).
The blue triangles indicate the positions of the five known X-ray sources\citep{2009ApJ...692..584H}, and the red pentagrams indicate the positions of the 13 known radio pulsars (M15A to O, except for M15I and N) in M15.
M15I\citep{2021ApJ...915L..28P} and N have been excluded because they have no unique phase-connected timing solutions.
The timing solution of M15M will be in our future paper (Dai et al. 2025 in preparation). 
The core radius (0.14$'$) and half-light radius (1$'$) are shown in dashed red circle and dashed blue circle, respectively.
The black cross in the right plot denotes the central position of M15.}
\label{Fig4}
\end{figure}

\begin{table}[h]  
\centering  
\caption{Timing parameters for M15O (PSR J2129-1210O), 
obtained from fitting the observed ToAs with \textsc{TEMPO}.
The DE421 Solar System Ephemeris and the Barycentric Dynamic Time time units are used, 
and the times were referenced to UTC(NIST).
}  
\begin{tabular}{lc}  
\hline\hline  
Pulsar\dotfill  & \textbf{J2129-1210O}  \\
\hline  
Right Ascension, $\alpha$ (J2000)\dotfill  & 21:29:58.2999(9)  \\
\hline  
Declination, $\delta$ (J2000)\dotfill  & +12:10:00.97(2) \\
\hline  
Spin Frequency, $f$ (s$^{-1}$)\dotfill & 90.3597874705(1)\\
\hline  
First Spin Frequency derivative, $\dot{f}$ (Hz s$^{-1}$)\dotfill & 1.79191(5)$\times 10^{-14}$   \\
\hline  
Second Spin Frequency derivative, $\ddot{f}$ (Hz s$^{-2}$)\dotfill & 3.3133(6)$\times 10^{-23}$   \\
\hline  
Reference Epoch (MJD)\dotfill & 60297.270620 \\
\hline  
Start of Timing Data (MJD)\dotfill  & 58404.483  \\
\hline  
End of Timing Data (MJD)\dotfill  & 60357.276\\
\hline  
Dispersion Measure, DM (pc cm$^{-3}$)\dotfill  & 67.44(1) \\
\hline  
Number of TOAs\dotfill    & 65 \\
\hline  
Residuals RMS ($\mu$s) \dotfill  & 108.00  \\
\hline\hline  
\end{tabular}  
\label{Tab2}
\end{table}
The timing analysis was performed with the TEMPO package.
We employed the \texttt{Dracula}\footnote{\url{https://github.com/pfreire163/Dracula}} algorithm \citep{2018MNRAS.476.4794F} to determine the precise rotation counts of pulsars.
The solution (see Table~\ref{Tab2}) returned from \texttt{Dracula} was consistent with the results obtained by manually removing the arbitrary phase between epochs.
The pulse profile and timing residuals of M15O are presented in Figure~\ref{Fig3}.
The position of M15O has a projected distance of 0.37$''$ from the cluster center and 0.81$''$ from M15A (Figure~\ref{Fig4}), making it the closest known pulsar to the M15 core in the projected sky. 
This confirms that M15O and M15A are spatially distinct.


\section{Discussion}
\label{discussion}

\subsection{Possible Reasons of Missing M15O in Previous Data Reduction}

We suggest that the key point in M15O identification is from the spectra.
As the peak near the 10$^{th}$ harmonic of M15A was seen, it is clear that such a signal can be something different from the harmonic itself.
The M15O is somehow easy to be detected by \textsc{PRESTO} based on typical pulsar search pipeline \citep{2020RAA....20...91Y}. 
We looked back to the previous pulsar search results and found that M15O should be discovered in January of 2021 or even earlier.
There are two possible reasons as bellows.

Firstly, periodic proximity introduces selection biases in candidate evaluation.
The harmonics of M15A can be the main affection in the identification of M15O.
The discovery plot (Figure ~\ref{Fig2}, left) is not so useful in this situation that the M15A signal dominates the features in all the panels.
A quick test showed that 
at least 50\% of experienced persons (colleagues around us, including experienced persons who found numbers of pulsars) were not agree that a new pulsar signal can be see in such a plot. 
Crucially, definitive confirmation of M15O required phase coherent timing analysis (position differences) rather than conventional a re-detection, 
indicating that this signal is quite easy to be mixed with the harmonics of M15A. 

Secondly, M15O's intrinsic faintness presents detection challenges even in FAST data. 
Its SNR is around 11 in the power spectrum, similar to the pulsar M15M and M15N which were discovered by stacking $\sim$44 hours FAST observational data. 
Thanks to M15O's relatively narrow pulse profile, making it easily to be distinguished from the noise level.

The radiometer equation is widely used to calculate the minimum detectable flux density of a pulsar (\citealt{2012hpa..book.....L}), as expressed by: \begin{equation} S_{\rm min} = \beta \frac{(S/N_{\rm min}) T_{\rm sys}}{G \sqrt{n_{\rm p} T_{\rm obs} \Delta f}} \sqrt{\frac{W_{\rm obs}}{P_{\rm spin} - W_{\rm obs}}}, \end{equation} where $\beta \sim 1$ represents the sampling efficiency of the 8-bit recording system. 
The system temperature $T{\rm sys}$ is 20~K, the antenna gain $G$ is 16\,K\,Jy$^{-1}$, and the number of polarizations $n_{\rm p}$ is 2. The bandwidth $\Delta f$ is set to 400 MHz \citep{2020RAA....20...64J}.
The ${W_{\rm obs}}$ is 0.15$P_{\rm spin}$ for M15O.
The estimated flux densities of M15O range from 1.02 to 2.72 $\mu\mathrm{Jy}$, as shown in Table~\ref{Tab1}. 

\subsection{The Large F1 and Measured F2 of M15O}

A phase connected timing solution of M15O could not be achieved by fitting only the spin frequency (F0) and first spin frequency derivative (F1). 
However, the inclusion of the second spin frequency derivative (F2) enabled us to obtain a single solution that connects all the available data (see Table~\ref{Tab2}).
A similar situation is observed for M15I.
It has a large F1 and also requires the inclusion of F2, but no single timing solution has been obtained so far.
Based on the timing results of M15O, the F1 
is remarkably large, second only to that of M15D among all known pulsars in M15 (see Table~\ref{Tab3}).
Moreover, M15O is the only pulsar in this cluster for which the F2 has been successfully measured.
M15O is located extremely close to the cluster center, with a projected distance of \(0.37''\).
These variations are likely caused by the complex gravitational potential near the center of M15, 
which may be due to the high-density interstellar environment or the orbital motion around a massive object. 
If the latter is the case, we may estimate the orbital period to be on the order of decades based on F1 and F2. 
Our observations with FAST have already spanned six years, and we plan to continue monitoring M15 with FAST. 
It is possible that future observations may reveal the nature of the very large F2 of either M15O or M15I.

\begin{table}[h]  
\centering  
\caption{Measured spin frequency and first spin frequency derivative parameters for the known pulsars (M15A-L) in M15 \citep{2024ApJ...974L..23W}. Timing for M15I and M15N was not obtained, marked with *. M15C is the only known binary, which is a pulsar-neutron star system with an orbital period of $\sim$0.34 days.} 
\begin{tabular}{lccc}  
\hline\hline
\textbf{Pulsar} & \textbf{Spin Frequency (s$^{-1}$)} & \textbf{F1 (Hz s$^{-1}$)} & \textbf{F2 (Hz s$^{-2}$)} \\
\hline
J2129+1210O & 90.3597874705(1) & $1.79191(5) \times 10^{-14}$ & 3.3133(6)$\times 10^{-23}$  \\ \hline
J2129+1210A & 9.0363060726718(5) & $1.713989(9) \times 10^{-15}$ & ...... \\
J2129+1210B & 17.814819156216(5) & $-3.02952(6) \times 10^{-15}$ & ...... \\
J2129+1210C & 32.7554227003(4) & $-5.3523(5) \times 10^{-15}$ & ...... \\
J2129+1210D & 208.21173107090(3) & $4.6391(4) \times 10^{-14}$ & ...... \\
J2129+1210E & 214.987399714593(8) & $-8.5721(1) \times 10^{-15}$ & ......\\
J2129+1210F & 248.32119058101(2) & $-1.6719(4) \times 10^{-15}$ & ......\\
J2129+1210G & 26.553254549(6) & $-1.152(6) \times 10^{-15}$ & ...... \\
J2129+1210H & 148.2932725188(7) & $-4.702(7) \times 10^{-16}$ & ...... \\
J2129+1210I* & 195.228710(2) & ...... & ...... \\
J2129+1210J & 84.44174596102(2) & $-1.4551(5) \times 10^{-15}$ & ...... \\
J2129+1210K & 0.51855097775(1) & $-3.174(2) \times 10^{-16}$ & ...... \\
J2129+1210L & 0.25247957713(1) & $-5.57(2) \times 10^{-17}$ & ...... \\
J2129+1210M & 206.77469911787(9) & $1.0127(1) \times 10^{-14}$ & ...... \\
J2129+1210N* & 107.64786(1) & ...... & ...... \\
\hline
\end{tabular}  
\label{Tab3}
\end{table}
\subsection{The Possibility of New Pulsars Hidden in the Known Pulsar Signals}

Pulsars observed in the same GC generally are located near the center and exhibit similar DM values. 
We conducted a statistical analysis of pulsars ($\sim 3600$ ones) outside of GCs from the PSRCAT\footnote{\url{https://www.atnf.csiro.au/research/pulsar/psrcat/}},
to see if any pair of pulsars can be included in a 5$'$ diameter circle and the DM difference between them is less than 5 \,cm$^{-3}$\,pc.
The results indicate that no such combinations were found. 
So, it seems that currently such hidden pulsar may not be missed in the Galactic field pulsar surveys.
On the other hand, it is worth noting that due to FAST's sensitivity, even more hidden pulsars may occur in GC data, even though M15O is the only case until now.

As radio telescope sensitivity has gradually increased in the past ten years, 
the number of detected pulsars keeps rising.
The pulsars discovered by FAST exceeded 1000 at the end of 2024, futher increasing the pulsar density in the Galactic filed.
The probability of encountering such cases, whether in GCs or on the Galactic disk, is also increasing.
We suggest that the spectra obtained from the FFT should be presented in a diagnostic plot. 
This will assist the identification of pulsar harmonics or new potential signals. 
We are looking forward to seeing if more can be identified through spectra-stacking searches on other GC data from FAST. 
In GC pulsars, we are checking the possible new pulsars hidden in the known pulsars harmonics, 
and will report the results in an upcoming paper focusing on the results from other GCs.

\section{Conclusions}
\label{conclusion}

The conclusions are as follows:

(1) An isolated MSP, namely M15O (J2129-1210O), 
with a period of $\sim$11.06687~ms and a DM of 67.44 \,cm$^{-3}$\,pc was discovered. 
Its spin period is very close to the 10th harmonic of the bright and previously known pulsar M15A ($\sim$11.06647 ms). 

(2) The discovery of M15O is due to the stack search method. 
To obtain pulsar candidates, each peak in the stacked spectrum was checked carefully.
The timing analysis was used to confirm this new pulsar.
This discovery validates the effectiveness of the stack search and improved candidate screening strategies in mitigating harmonic interference.

(3) We obtained the phase connected timing solution for M15O.
It has large F1 and measurable F2, being \(1.79191(5) \times 10^{-14}\) Hz s\(^{-1}\) and \(3.3133(6) \times 10^{-23}\) Hz s\(^{-2}\), respectively.
Its projected distance from the optical center of M15 is the closest among all the known pulsars in M15. 

(4) We suggest adding the power spectrum in the pulsar candidate diagnostic plot to identify possible new signals near the harmonics of previously known pulsars.

\begin{acknowledgements}

We thank the referee's valuable suggestions, which helped to refine our paper. 
We thank Paulo C. C. Freire and Scott M. Ransom for valuable discussions and suggestions during the validation of M15O, and Paulo C. C. Freire for helping us obtain the ephemeris.
This work is supported by the National Key R\&D Program of China No. 2022YFC2205202, No. 2020SKA0120100 and the National Natural Science Foundation of China (NSFC, Grant Nos. 12373032, 12003047, 11773041, U2031119, 12173052, and 12173053.
Both Lei Qian and Zhichen Pan were supported by the Youth Innovation Promotion Association of CAS (id.~2018075, Y2022027 and 2023064), and the CAS "Light of West China" Program.
Liyun Zhang has been supported by the Science and Technology Program of Guizhou Province under project No.QKHPTRC-ZDSYS[2023]003 and QKHFQ[2023]003. 
This work made use of the data from FAST (Five-hundred-meter Aperture Spherical radio Telescope) (https://cstr.cn/31116.02.FAST). 
FAST is a Chinese national mega-science facility, operated by National Astronomical Observatories, Chinese Academy of Sciences.

\end{acknowledgements}

\label{lastpage}
\end{document}